\documentclass[showpacs,preprintnumbers,amsmath,amssymb,prb,floatfix]{revtex4}
\usepackage{graphicx}
\usepackage{dcolumn}
\usepackage{bm}
\usepackage{amsmath,amsthm,amssymb,ascmac}

\pacs{}

\newcommand{\be}{\beta}

\newcommand{\bea}{\begin{eqnarray}}
\newcommand{\eea}{\end{eqnarray}}
\newcommand{\aeq}{\!\! &=& \!\!}

\newcommand{\aeqd}{\!\! & \stackrel{\mathrm{def}}{=} & \!\!}

\newcommand{\mL}{\mathcal{L}}

\newcommand{\mH}{\mathcal{H}}

\newcommand{\D}{{\rm{D}}}

\newcommand{\dg}{^\dagger}

\newcommand{\defe}{\stackrel{\mathrm{def}}{=}}

\newcommand{\ga}{\gamma}

\newcommand{\al}{\alpha}

\newcommand{\f}{\frac}
\newcommand{\half}{\frac{1}{2}}
\newcommand{\pr}{\prime}
\newcommand{\dl}{\delta}

\newcommand{\lm}{\lambda}

\newcommand{\om}{\omega}
\newcommand{\Om}{\Omega}

\newcommand{\tot}{{\rm{tot}}}

\newcommand{\la}{\label}

\newcommand{\no}{\nonumber}

\newcommand{\re}[1]{(\ref{#1})}

\newcommand{\res}[1]{˜\ref{#1}}

\newcommand{\Bv}[1]{\Big \vert_{#1}}

\newcommand{\RM}[1]{{\rm{#1}}}
\newcommand{\p}{\partial}
\newcommand{\w}{\wedge}
\newcommand{\co}[1]{``{#1}''}
\newcommand{\Ref}[1]{Ref.[\onlinecite{#1}]}
\newcommand{\DW}{{\rm{DW}}}

\begin{document}

\title{Reconsideration of De Donder-Weyl theory by covariant analytic mechanics}

\author{Satoshi Nakajima}
\email{subarusatosi@gmail.com}

\date{\today}

\affiliation{
Graduate School of Pure and Applied Sciences, University of Tsukuba, 
1-1-1, Tennodai, Tsukuba, Japan 305-8571}
\date{\today}

\begin{abstract}
We show that the covariant analytic mechanics (CAM) is closely related to the De Donder-Weyl (DW) theory.
To treat space and time on an equal footing,  
the DW theory introduces $D$ conjugate fields ($D$ is the dimension of space-time) for each field and the CAM regards the differential forms as the basic variables. 
The generalization of the canonical equations is called the DW equations. 
Although one of the DW equations is not correct for the gauge field and the gravitational field, we show the way to improve it.
By rewriting the canonical equations of the CAM, which are manifestly general coordinate covariant and gauge covariant, using the components of the tensors,
we show that these are equivalent to the improved DW equations. 
Additionally, we investigate the Dirac field. 
We present a modified Hamilton formalism which regards only the Dirac fields as the basic variables and show that it provides the Dirac equations correctly. 
\end{abstract}

\maketitle

\section{Introduction}

In the traditional analytic mechanics, the Hamilton formalism gives especial weight to time, then, the covariance is not trivial.
To solve this problem, the De Donder-Weyl (DW) theory\cite{Weyl,IV,IV_D} introduced the conjugate fields $\pi_a^\mu=\p \mL/\p \p_\mu \psi^a$ $(\mu=0,1,\cdots,D-1)$ for each 
field $\psi^a$. 
Here, $D$ is the dimension of space-time and $\mL$ is the Lagrangian density. 
The generalization of the canonical equations is called the DW equations.
If $a$ includes the tenser index, one  of the DW equations is not correct generally.

In the traditional analytic mechanics and the DW theory, the basic variables are the components of the tensors.
In contrast, the covariant analytic mechanics (CAM) regards the differential forms as the basic variables. 
Because the differential form is independent of the coordinate system, the general coordinate covariance is guaranteed manifestly.
The Lagrange formalism of the CAM has been used \cite{75K,75T,80,95} from a few decades before.
Nakamura \cite{Na} generalized this formulation to the Hamilton formalism. 
In this method, the conjugate form is also a differential form, which treats space and time on an equal footing. 
Nakamura applied this method to the Proca field and the electromagnetic field with manifest covariance and, in the latter, with the gauge covariance. 
Kaminaga \cite{K} formulated strictly mathematically Nakamura's idea and constructed the general theory in arbitrary dimension. 
Kaminaga studied that the Newtonian mechanics of a harmonic oscillator and the scalar field, 
 the electromagnetic field, the non-abelian gauge fields and 4 dimension gravity (second order formalism) without the Dirac field. 
On the other hand, Nester \cite{N04} also independently investigated the CAM and 
 constructed the general theory in 4 dimension and applied it to the Proca field, the electromagnetic field and the non-abelian gauge field.
However, the treatment of the gravitational field (first order formalism) was not complete Hamilton formalism because the quantity corresponding to 
the Hamiltonian density was given by hand \cite{N91}. 
We applied the CAM to the gravity with the Dirac field (both first and second order formalism)\cite{Me}. 

In this manuscript, we improve one of the DW equation and show that the improved DW theory is equivalent to the CAM for the first time. 
We apply the CAM to the Dirac field with the Lagrange multiplier.
Moreover, we present a modified Hamilton formalism which regards only the Dirac fields as the basic variables and show that it provides the Dirac equations correctly. 

In \S 2, we discuss the relation between the DW theory and the CAM. 
We introduce the DW theory in \res{DWT} and the CAM in \res{CAM}. 
In \res{Relation}, we introduce improved DW equations and show that these are equivalent to the canonical equations of the CAM by rewriting the latter using the components of the tensors. 
And we discuss the difference between the Poisson bracket of the CAM  and that of the DW theory\cite{IV,IV_D}.
In \S 3, we apply the CAM to the Dirac field, which is a constraint system.
We treat the Dirac field with the Lagrange multiplier.
Moreover, using a modified Hamilton formalism regarding only the Dirac fields as the basic variables, we derive the Dirac equations correctly.

\section{De Donder-Weyl theory and covariant analytic mechanics} 

\subsection{De Donder-Weyl theory} \la{DWT}

Let us consider $D$ dimension space-time. 
The Lagrangian density is a function of fields $\psi_{\mu_1 \cdots \mu_p}^A$ and $\partial_\mu \psi_{\mu_1 \cdots \mu_p}^A$F
$\mL=\mL(\psi_{\mu_1 \cdots \mu_p}^A,\partial_\mu \psi_{\mu_1 \cdots \mu_p}^A)$. 
Here, $A$ does not include tensor index and $p$ $(=0,1,\cdots)$ can depend on $A$. 
The Euler-Lagrange equation is given by
\bea
\f{\partial \bm{\mL}}{\partial \psi_{\mu_1 \cdots \mu_p}^A}-\partial_\mu\f{\partial \bm{\mL}}{\partial \partial_\mu \psi_{\mu_1 \cdots \mu_p}^A} \aeq 0, \la{EL_u}
\eea
where $\bm{\mL}=\sqrt{-g}\mL$ and $g$ is the determinant of the metric $g_{\mu\nu}$. The conjugate fields of $\psi_{\mu_1 \cdots \mu_p}^A$ are defined by
\bea
\bm{\pi}_A^{\mu,\mu_1\cdots \mu_p} \aeqd \f{\partial \bm{\mL}}{\partial \partial_\mu \psi_{\mu_1 \cdots \mu_p}^A} ,\la{pi_DW}
\eea
and $\pi_A^{\mu,\mu_1\cdots \mu_p}\defe \bm{\pi}_A^{\mu,\mu_1\cdots \mu_p}/\sqrt{-g}$.
These are also called the generalized momenta or polymomenta. 
$\pi_A^{0,\mu_1\cdots \mu_p}$ is the traditional conjugate momentum. 
The DW Hamiltonian density is defined by
\bea
\bm{\mH}_\DW(\psi_{\mu_1 \cdots \mu_p}^A,\bm{\pi}_A^{\mu,\mu_1\cdots \mu_p}) \aeqd \partial_\mu \psi_{\mu_1 \cdots \mu_p}^A\bm{\pi}_A^{\mu,\mu_1\cdots \mu_p}-\bm{\mL}, \la{H_DW}
\eea
and $\mH_\DW\defe\bm{\mH}_\DW/\sqrt{-g}$. The variation of $\bm{\mH}_\DW$ is given by
\bea
\dl \bm{\mH}_\DW 
\aeq \partial_\mu \psi_{\mu_1 \cdots \mu_p}^A \dl \bm{\pi}_A^{\mu,\mu_1\cdots \mu_p}-\f{\partial \bm{\mL}}{\partial \psi_{\mu_1 \cdots \mu_p}^A} \dl \psi_{\mu_1 \cdots \mu_p}^A .
\la{dl_H_DW}
\eea
If $\bm{\pi}_A^{0,\mu_1\cdots \mu_p},\cdots,\bm{\pi}_A^{D-1,\mu_1\cdots \mu_p}$ are independent with each other (this assumption is not correct generally) 
and these are independent of $\psi_{\mu_1 \cdots \mu_p}^A$, we obtain identities
\bea
\f{\partial \bm{\mH}_\DW}{\partial \psi_{\mu_1 \cdots \mu_p}^A} \aeq -\f{\partial \bm{\mL}}{\partial \psi_{\mu_1 \cdots \mu_p}^A} ,\
\f{\partial \bm{\mH}_\DW}{\partial \bm{\pi}_A^{\mu,\mu_1\cdots \mu_p}} = \partial_\mu \psi_{\mu_1 \cdots \mu_p}^A. \la{I_DW}
\eea
Substituting \re{EL_u} to the first equation, we obtain the DW equations:
\bea
\partial_\mu \psi_{\mu_1 \cdots \mu_p}^A \aeq \f{\partial \bm{\mH}_\DW}{\partial \bm{\pi}_A^{\mu,\mu_1\cdots \mu_p}} , \la{DW1}\\
\partial_\mu \bm{\pi}_A^{\mu,\mu_1\cdots \mu_p} \aeq -\f{\partial \bm{\mH}_\DW}{\partial \psi_{\mu_1 \cdots \mu_p}^A} .\la{DW2}
\eea

For instance, the electromagnetic field corresponds to $p=1$. 
The Lagrangian density is given by $\mL(A_\nu,\partial_\mu A_\nu)=-\f{1}{4}F_{\mu\nu}F^{\mu\nu}+A_\nu J^\nu$ with
$F_{\mu\nu}=2\partial_{[\mu}A_{\nu]}=\partial_{\mu}A_{\nu}-\partial_{\nu}A_{\mu}$. 
[ ] is the anti-symmetrization symbol. 
$A_\nu$ is the vector potential and $J^\nu$ is the current density, which is independent of $A_\mu$. 
All indices are lowered and raised with $g_{\mu\nu}$ or its inverse $g^{\mu\nu}$. 
The conjugate fields of $A_\nu$ are $\pi^{\mu,\nu}=-F^{\mu\nu}=-\pi^{\nu,\mu}$. 
The DW Hamiltonian density is 
$\mH_\DW(A_\nu,\pi^{\mu,\nu})=-\f{1}{4}\pi_{\mu,\nu}\pi^{\mu,\nu}-A_\nu J^\nu$. 
We have $\partial \bm{\mH}_\DW/\partial \bm{\pi}^{\mu,\nu}=-\half \pi_{\mu,\nu}$ and $\partial \bm{\mH}_\DW/\partial A_\nu=-\sqrt{-g}J^\nu $. 
Then, the DW equations are given by
\bea
\partial_\mu A_\nu =-\half \pi_{\mu,\nu} , \ \partial_\mu \bm{\pi}^{\mu,\nu}=\sqrt{-g}J^\nu.
\eea
The letter is equivalent to the Maxwell equation $ \partial_\mu(\sqrt{-g}F^{\mu\nu})=-\sqrt{-g}J^\nu$. 
However, the former is not correct. 
This is because $\pi^{\mu,\nu}$ are supposed as independent each other although there is a constraint condition $\pi^{\mu,\nu}=-\pi^{\nu,\mu}$.

\subsection{Covariant analytic mechanics} \la{CAM}

Suppose a $p$-form $\be$ ($p=0,1,\cdots,D$) is described by forms $\{\al^I \}_{I=1,\cdots,k}$. 
If there exists the form $\om_I$ such that $\be$ behaves under variations $\dl \al^I$ as 
\bea
\dl \be = \sum_I \dl \al^I \wedge \om_I ,
\eea
we call $\om_I$ the {\it derivative} of $\be$ by $\al^I$ and denote 
\bea
\f{\partial \be}{\partial \al^I} \defe  \om_I.
\eea
If $\al^I$ is $q_I(\le p)$-form, $\partial \be/\partial \al^I$ is $(p-q_I)$-form. 
Although the derivative of arbitrary form dose not always exist, as explained in \res{Relation}, the derivative of $D$-forms always exist.
The Hodge operator $\ast$ maps an arbitrary $p$-form $\om=\om_{\mu_1 \cdots \mu_p}dx^{\mu_1}\w \cdots \w dx^{\mu_p}$ $(p=0,1,\cdots,D)$ to a $D-p=r$-form as
\bea
\ast \om = \f{1}{r!}E_{\nu_1 \cdots \nu_r}^{\ \ \ \ \ \ \mu_1 \cdots \mu_p}\om_{\mu_1 \cdots \mu_p}dx^{\nu_1} \w \cdots \w dx^{\nu_r}.
\eea
Here, $\om=\om_{\mu_1 \cdots \mu_p}$ and $E_{\mu_1 \cdots \mu_D}$ are complete anti-symmetric tensors and $E_{01\cdots D-1}=\sqrt{-g}$. 
In particular, $\Om=\ast 1$ is the volume form. 
And $\ast \ast \om=-(-1)^{p(D-p)}\om$ holds.

As the traditional analytic mechanics and the DW theory start from the Lagrangian density, the CAM starts from the {\it Lagrange $D$-form} $L$ 
defined by $L \defe \mL \Om$. 
$L$ is assumed to be described by forms $\psi^A$ and $d\psi^A$: $L=L(\psi^A,d\psi^A)$. 
If $\psi^A$ is a $p$-from ($p=0,1,\cdots,D-1$), $\psi^A$ and $d\psi^A$ are expanded using a $p$-th order complete anti-symmetric tensor $\psi^A_{\mu_1 \cdots \mu_p}$ as 
$\psi^A=\psi^A_{\mu_1 \cdots \mu_p}dx^{\mu_1}\w \cdots \w dx^{\mu_p}$ and $d\psi^A=\partial_{[\mu} \psi^A_{\mu_1 \cdots \mu_p]}dx^\mu\w dx^{\mu_1}\w \cdots \w dx^{\mu_p}$. 
Namely, the CAM treats a special class of the field theories of which Lagrangian density is described by the complete anti-symmetric tensors $\psi^A_{[\mu_1 \cdots \mu_p]}$ 
and $\partial_{[\mu} \psi^A_{\mu_1 \cdots \mu_p]}$. 
All fundamental field theories belong to this class. 
For instance, the Lagrange form of the electromagnetic field is given by
\bea
L(A,dA)=-\half F\w \ast F+J\w A ,
\eea
where $A=A_\mu dx^\mu$, $J=\ast (J_\mu dx^\mu)$ and $F=dA=\half F_{\mu\nu}dx^\mu \w dx^\nu$.

The Euler-Lagrange equation is given by
\bea
\f{\partial L}{\partial \psi^A}-(-1)^p d\f{\partial L}{\partial d\psi^A} = 0. \la{EL}
\eea
The {\it conjugate form} $\pi_A$ is defined by
\bea
\pi_A \defe \f{\partial L}{\partial d\psi^A}.
\eea
$\pi_A$ is $D-p-1=q$-form. The {\it Hamilton $D$-form} (not $(D-1)$-form) is defined by
\bea
H(\psi^A,\pi_A)\defe d\psi^A \wedge \pi_A-L . \la{def_H}
\eea
The variation of $H$ is given by
\bea
\dl H = (-1)^{(p+1)q}\dl \pi_A \wedge d\psi^A -\dl \psi^A \wedge\f{\partial L}{\partial \psi^A}. \la{dl_H}
\eea
If $\psi^A$ and $\pi_A$ are independent, we get identities
\bea
\f{\partial H}{\partial \psi^A}=  -\f{\partial L}{\partial \psi^A},\
\f{\partial H}{\partial \pi_A}= (-1)^{(p+1)q}d\psi^A. \la{der_H}
\eea
Here, $(-1)^{(p+1)q}=1$ if $p$ is an odd number and $(-1)^{(p+1)q}=-(-1)^D$ if $p$ is an even number.
By substituting the Euler-Lagrange equation \re{EL} into the first equation of \re{der_H}, we obtain the {\it canonical equations} \cite{K,Me}
\bea
d\psi^A  \aeq (-1)^{(p+1)q}\f{\partial H}{\partial \pi_A} ,\la{C1}\\
 d\pi_A \aeq -(-1)^p\f{\partial H}{\partial \psi^A}.\la{C2}
\eea
The inverse transformation of \re{def_H} is given by 
\bea
L^\pr(\psi,\nu) \defe \pi_A \w \nu^A-H ,\ \nu^A \defe \f{\p H}{\p \pi_A}.
\eea
From the second equation of \re{der_H}, we have $\nu^A=(-1)^{(p+1)q}d\psi^A$. 
Then, $L^\pr(\psi,\nu)$ reduces to the Lagrange form: $L^\pr(\psi,\nu)=L(\psi,d\psi)$.

For the electromagnetic field, the Euler-Lagrange equation is given by $d\ast F=-(-1)^D J $.
The conjugate form, $\pi=\partial L/\partial dA=-\ast F$, can represent $dA$ as $dA=\ast \pi$.
The Hamilton form is given by
\bea
H(A,\pi) = \half \pi \w \ast \pi-J\w A.
\eea
We have $\partial H/\partial \pi=\ast \pi$ and $\partial H/\partial A=(-1)^DJ$. 
The canonical equations $dA=\partial H/\partial \pi$ and $d\pi=\partial H/\partial A$ are
\bea
dA=\ast \pi,\ d\pi=(-1)^D J .\la{Maxwell}
\eea
The former is equivalent to the definition of the conjugate form and the latter 
coincides with the Euler-Lagrange equation.

We introduce two {\it Poisson brackets} by
\bea
\{\al,\be\}_0
\aeqd (-1)^{(p+1)q}  \f{\partial \al}{\partial \psi^A}\w \f{\partial \be}{\partial \pi_A}-(-1)^p  \f{\partial \al}{\partial \pi_A} \w \f{\partial \be}{\partial \psi^A}, \la{PB0} \\
\{\al,\be\}_1 
\aeqd (-1)^{(p+1)q} \f{\partial \be}{\partial \pi_A}\w \f{\partial \al}{\partial \psi^A}-(-1)^p \f{\partial \be}{\partial \psi^A}\w \f{\partial \al}{\partial \pi_A}, \la{PB1}
\eea
for arbitrary forms $\al$ and $\be$ which have the derivatives by $\psi^A$ and $\pi_A$. 
\re{PB0} was introduced in \Ref{Me}. 
The canonical equations \re{C1} and \re{C2} can be written as
\bea
d\psi^A  =\{ \psi^A,H\}_i ,\ d\pi_A  =\{ \pi_A,H\}_i, \la{Cs}
\eea
for $i=0,1$. And we have  
\bea
\{\psi^A,\psi^B\}_i= 0,\ \{\pi_A,\pi_B\}_i \aeq0,\no\\
\{\psi^A,\pi_B\}_i= (-1)^{(p+1)q}\dl^A_B,\ \{\pi_B,\psi^A\}_i \aeq-(-1)^p\dl^A_B, \la{FPB}
\eea
for $i=0,1$. 
Moreover, the Leibniz rules
\bea
\{\al,\be\w \ga \}_0\aeq \{\al,\be\}_0\w \ga+(-1)^{b(D+1-a)}\be\w \{\al,\ga\}_0 ,\la{LR0} \\
\{\al\w \be,\ga \}_1\aeq \{\al,\ga\}_1\w \be+(-1)^{a(D+1-c)}\al\w \{\be,\ga \}_1 \la{LR1}
\eea
hold for arbitrary $a,b$ and $c$ forms $\al$, $\be$ and $\ga$ which have the derivatives by $\psi^A$ and $\pi_A$. 
Using \re{LR1} and \re{Cs}, the equation of motions of $\al \w \be=\psi^A\w \psi^B, \psi^A \w \pi_B, \pi_A \w \pi_B$ are given by
\bea
d(\al \w \be) \aeq \{ \al \w \be,H\}_1 . \la{new}
\eea
In the derivation of \re{LR0} and \re{LR1}, we used 
\bea
\f{\p (\al \w \be)}{\p \psi^A} =\f{\p \al}{\p \psi^A} \w \be+(-1)^{ap} \al \w \f{\p \be}{\p \psi^A},\
\f{\p (\al \w \be)}{\p \pi_A} =\f{\p \al}{\p \pi_A} \w \be+(-1)^{aq} \al \w \f{\p \be}{\p \pi_A}.
\eea
If the derivative of $\al$ or $\be$ by $\psi^A(\pi_A)$  does not exist, the right hind side of the above first (second) equation is meaningless. 
For instance, if $\al \w \be=\dl^{BC} \pi_B \w \ast \pi_C$, although $\p (\al \w \be)/\p \pi_A=2\ast\pi_A$ holds,
the above second equation does not hold since $\p \ast \pi_B/\p \pi_A$ does not exist (except for $D=1$).

\subsection{Relation between the two theories} \la{Relation}

In the CAM, the derivative of the fields is included as $\partial_{[\mu} \psi^A_{\mu_1 \cdots \mu_p]}$.
Then, the conjugate fields defined by \re{pi_DW} satisfy 
\bea
\bm{\pi}_A^{\mu,\mu_1\cdots \mu_p}=\bm{\pi}_A^{[\mu,\mu_1\cdots \mu_p]}. \la{con}
\eea
Considering this condition at \re{dl_H_DW}, we obtain an improved equation of the second equation of \re{I_DW} and of the DW equation \re{DW1}:
\bea
\partial_{[\mu} \psi_{\mu_1 \cdots \mu_p]}^A \aeq \f{\partial \bm{\mH}_\DW}{\partial \bm{\pi}_A^{\mu,\mu_1\cdots \mu_p}}. \la{DW1a}
\eea
For the electromagnetic field, this equation becomes $\partial_{[\mu} A_{\nu]} =-\half \pi_{\mu,\nu} $. 
This is correct and corresponds to the first equation of \re{Maxwell}.
In the following, we show that \re{C1} and \re{C2} are respectively equivalent to \re{DW1a} and \re{DW2}. 

We put $\ast \partial L/\partial \psi^A=m_{A\mu_1\cdots\mu_p}dx^{\mu_1}\w \cdots \w dx^{\mu_p}$. 
From $\dl \psi^A \w \partial L/\partial \psi^A=-\ast(\ast \partial L/\partial \psi^A)\w \dl \psi^A$, 
we have 
\bea
\dl \psi^A \w \f{\partial L}{\partial \psi^A}= -p!m_A^{\mu_1\cdots\mu_p}\dl \psi^A_{\mu_1\cdots\mu_p} \Om .
\eea
On the other hand, $\dl L$ is given by
\bea
\dl L=\Big[\f{\partial \bm{\mL}}{\partial \psi^A_{\mu_1\cdots\mu_p}}\dl \psi^A_{\mu_1\cdots\mu_p}
+ \bm{\pi}_A^{\mu,\mu_1\cdots\mu_p}\dl \partial_{[\mu}\psi^A_{\mu_1\cdots\mu_p]}\Big]\f{\Om}{\sqrt{-g}} .
\eea
Then, we get $ m_A^{\mu_1 \cdots \mu_p} =-\f{1}{p!} \f{1}{\sqrt{-g}}\f{\partial \bm{\mL}}{\partial \psi^A_{\mu_1 \cdots \mu_p}}$. 
It leads 
\bea
 \f{\partial L}{\partial \psi^A} \aeq \f{(-1)^{p(D-p)}}{p!} \f{1}{\sqrt{-g}}\f{\partial \bm{\mL}}{\partial \psi^A_{\mu_1 \cdots \mu_p}}\ast dx_{\mu_1} \w \cdots \w dx_{\mu_p}, \la{com1}
\eea
with $dx_\mu=g_{\mu\nu}dx^\nu$. 
Similarly, we obtain 
\bea
\pi_A  \aeq  \f{(-1)^{(p+1)q}}{(p+1)!}\pi_A^{\mu,\mu_1 \cdots \mu_p}\ast dx_\mu \w dx_{\mu_1} \w \cdots \w dx_{\mu_p} \no\\
\aeq  \f{(-1)^{(p+1)q}}{(p+1)!q!}\bm{\pi}_A^{\mu,\mu_1 \cdots \mu_p}\f{1}{\sqrt{-g}}E_{\nu_1 \cdots \nu_q,\mu\mu_1 \cdots \mu_p}dx^{\nu_1 } \w \cdots \w dx^{\nu_q}. \la{com2}
\eea
It leads
\bea
d\pi_A  \aeq  \f{(-1)^{(p+1)q}}{(p+1)!q!}\partial_\lm\bm{\pi}_A^{\mu,\mu_1 \cdots \mu_p}\f{1}{\sqrt{-g}}E_{\nu_1 \cdots \nu_q,\mu\mu_1 \cdots \mu_p}
dx^\lm\w dx^{\nu_1 } \w \cdots \w dx^{\nu_q}.
\eea
From this equation, we obtain 
\bea
\ast d\pi_A \aeq  -\f{(-1)^p}{p!}\f{1}{\sqrt{-g}}\partial_\mu\bm{\pi}_A^{\mu,\mu_1 \cdots \mu_p}dx_{\mu_1} \w  \cdots \w dx_{\mu_p} , 
\eea
using a formula $E_{\nu_1 \cdots \nu_a\mu_1 \cdots \mu_b}E^{\nu_1 \cdots \nu_a\rho_1 \cdots \rho_b} =-a!b!\dl_{[\mu_1}^{\rho_1} \cdots \dl_{\mu_b]}^{\rho_b} \ (a+b=D)$. 
The above equation is equivalent to 
\bea
d\pi_A \aeq  \f{(-1)^{pq}}{p!}\f{1}{\sqrt{-g}}\partial_\mu\bm{\pi}_A^{\mu,\mu_1 \cdots \mu_p}\ast dx_{\mu_1} \w  \cdots \w dx_{\mu_p}. \la{com3}
\eea
Because of \re{com1} and \re{com3}, the Euler-Lagrange equation \re{EL} is equivalent to \re{EL_u}.
By the way, 
$d\psi^A \w \pi_A=\partial_{[\mu} \psi^A_{\mu_1 \cdots \mu_p]}\pi_A^{\mu,\mu_1 \cdots \mu_p} \Om= \partial_{\mu} \psi^A_{\mu_1 \cdots \mu_p} \pi_A^{\mu,\mu_1 \cdots \mu_p}\Om$ 
holds. 
It leads 
\bea
H=\mH_\DW \Om=\bm{\mH_\DW}\f{\Om}{\sqrt{-g}} .
\eea
Using this equation, we get an equation similar to \re{com1}:
\bea
 \f{\partial H}{\partial \psi^A} \aeq \f{(-1)^{p(D-p)}}{p!} \f{1}{\sqrt{-g}}\f{\partial \bm{\mH}_\DW}{\partial \psi^A_{\mu_1 \cdots \mu_p}}\ast dx_{\mu_1} \w \cdots \w dx_{\mu_p}.
\eea
Because of the above equation and \re{com3}, the canonical equation \re{C2} is equivalent to \re{DW2}. 
If we put $\partial H/\partial \pi_A=l^A_{\mu\mu_1\cdots\mu_p}dx^\mu \w dx^{\mu_1}\w \cdots \w dx^{\mu_p}$, we have
\bea
\dl \pi_A \w \f{\partial H}{\partial \pi_A}=(-1)^{(p+1)q}\dl \bm{\pi}_A^{\mu,\mu_1 \cdots \mu_p}l_{\mu\mu_1 \cdots \mu_p}^A\f{\Om}{\sqrt{-g}}.
\eea
It means $l^A_{\mu\mu_1\cdots\mu_p}=(-1)^{(p+1)q}\partial \bm{\mH}_\DW/\partial \bm{\pi}_A^{\mu,\mu_1 \cdots \mu_p}$. 
Then, we obtain
\bea
\f{\partial H}{\partial \pi_A} \aeq (-1)^{(p+1)q}\f{\partial \bm{\mH}_\DW}{\partial \bm{\pi}_A^{\mu,\mu_1 \cdots \mu_p}}dx^\mu \w dx^{\mu_1}\w \cdots dx^{\mu_p} .
\eea
Because of this equation, the canonical equation \re{C1} becomes \re{DW1a}.

In the DW theory, the Poisson-Gerstenhaber bracket (PGB), which is similar with \re{PB0} and \re{PB1}, had been introduced 
using forms and multivector fields on the extended polymomentum phase space\cite{IV} of which coordinate is $(\psi^a,\pi^\mu_a,x^\mu)$. 
Here, $a$ denotes both  $A$ and $\mu_1 \cdots \mu_p$. 
While our Poisson brackets are defined using the forms on space-times, the PGB is defined using the components of the tensors. 
Because the constraint \re{con} is not considered, the PGB does describe incorrect equation \re{DW1}. 
Because of use of the the extended polymomentum phase space, the PGB has abundant structure and the scope of application of the PGB is wider than \re{PB0} and \re{PB1}. 
The PBG satisfies only the graded Leibniz rule but also the graded anticommutativity and the graded Jacobi identity. 
While a generalization of the PGB to the Dirac bracket had been studied \cite{IV_D}, the generalization of our Poisson bracket is not known yet.

In the DW thoery, \re{DW1} is incorrect since the constraint \re{con} is not considered. 
However, decreasing independent variables using \re{con}, we obtain correct equation \re{DW1a}. 
This fact suggests that one can obtain correct equation of motions from the Hamilton form by the variation using only independent variables 
determined by constraints among the original variables $\psi^A_{\mu_1\cdots \mu_p}$ and $\bm{\pi}_A^{[\mu,\mu_1\cdots \mu_p]}$. 
We confirm this expectation for the Dirac field in the next section.

\section{Dirac field} \la{Dirac field}

To obtain \re{der_H} from \re{dl_H}, we supposed that $\psi^A$ and $\pi_A$ are independent. 
However, if $\bm{\pi}_A^{\mu,\mu_1 \cdots \mu_p}$ is a function of $\psi^A_{\mu_1 \cdots \mu_p}$ or identically zero, 
 $\pi_A$ is not an independent variable.
The Dirac field and the first order formalism of the gravitational field correspond to this case. 
In this section, we study the Dirac field based on the CAM. 
First we use the method of Lagrange multiplier, which is the standard method for constraint systems. 
And next, we present a modified Hamilton formalism regarding only the Dirac fields as the basic variables 
to confirm the expectation mentioned in \res{Relation}. 
In \Ref{GS}, the Dirac field had been studied based on the DW theory without using the Lagrange multiplier.

In the following of this paper, we set $D=4$.  Let $\{\theta^a\}_{a=0,1,2,3}$ denote an orthonormal frame.
$\theta^a$ can be expanded as $\theta^a=\theta^a_{\ \mu}dx^\mu$ with the vielbein $\theta^a_{\ \mu}$. 
We have $g_{\mu\nu}=\eta_{ab}\theta^a_{\ \mu} \theta^b_{\ \nu}$ with $\eta_{ab}=\RM{diag}(-+++)$. 
All indices are lowered and raised with $\eta_{ab}$ or its inverse $\eta^{ab}$.
Let $\om^a_{\ b}$ be the connection 1-form. 
$\om_{ba}=-\om_{ab}$ holds. 
The torsion 2-form $\Theta^a=\half C^a_{\ bc}\theta^b \w \theta^c$ satisfies 
$d\theta^a+\om^a_{\ b}\w \theta^b=\Theta^a $.

The Lagrange form of the Dirac field $\psi={}^t(\psi^1,\psi^2,\psi^3,\psi^4)$ is given by
\bea
L_\D^\be
 \aeq -\f{1+\be}{2} \bar{\psi}\ga_c e^c \w (d\psi+\f{1}{4}\ga_{ab}\om^{ab}\psi)+\f{1-\be}{2} e^c \w (d\bar{\psi}-\f{1}{4}\bar{\psi}\ga_{ab}\om^{ab})\ga_c\psi
-m\bar{\psi}\psi \Om  \no\\
&&-\f{\be}{2}C_a\bar{\psi}\ga^a\psi \Om ,
\eea
with $\bar{\psi}\defe i\psi \dg \ga^0=(\bar{\psi}_1,\bar{\psi}_2,\bar{\psi}_3,\bar{\psi}_4)$, $e^a \defe \ast \theta^a$, $C_a\defe C^b_{\ ab}$ and $\ga_{ab}\defe \ga_{[a}\ga_{b]}$. 
Here, $\ga^a$ is the gamma matrix, which satisfies $\ga^{a}\ga^{b}+\ga^{b}\ga^{a}=2\eta^{ab}$. 
$\be$ is an arbitrary real number. $m$ is the mass.
$L_\D^\be$ is given by $L_\D^{\be=0}+\f{\be}{2} d(e^a\bar{\psi}\ga_a\psi)$ because of 
\bea
d e_a = -(\om_a+C_a)\Om ,\
 \ga_c  e^c \w \f{1}{4}\ga_{ab}\om^{ab}=\f{1}{4}  e^c \w \ga_{ab}\om^{ab}\ga_c+\ga^a \om_a \Om, \la{trick}
\eea
with $\om_a \defe \om^b_{\ ab}$.  
Because the Dirac field is a representation of the Lorentz transformation and 
a spinor field is defined in the tangent Minkowski space, the frame is necessary to write down the Lagrange form even in the flat space-time.
The connection $\om^a_{\ b}$ is the gauge field for the local Lorentz transformations. 
$C_a$ is regarded as independent of $\psi$ and $\bar{\psi}$. 
For simplicity, we treat the Dirac field as a usual number (not the Grassmann number).
The Euler-Lagrange equations $\partial L_\D^\be/\partial \psi^A-d(\partial L_\D^\be/\partial d\psi^A)=0$  
and $\partial L_\D^\be/\partial \bar{\psi}_A-d(\partial L_\D^\be/\partial d\bar{\psi}_A)=0$ ($A=1,2,3,4$)
are respectively given by
\bea
(d\bar{\psi} -\f{1}{4} \bar{\psi}\ga_{ab} \om^{ab})\w \ga_c e^c+m\bar{\psi}\Om -\half C_a\bar{\psi}\ga^a  \Om \aeq 0,\la{DE2} \\
\ga_c e^c \w (d+\f{1}{4}\ga_{ab}\om^{ab})\psi+m \psi\Om+\half C_a \ga^a\psi \Om \aeq 0 ,\la{DE} 
\eea
using \re{trick} \cite{Me}. \re{DE2} is the Hermitian conjugate of \re{DE}.

The conjugate forms of $\psi^A$ and $\bar{\psi}_A$ are given by
\bea
\Pi^\be_A=\f{\partial L_\D^\be}{\partial d\psi^A}=\f{1+\be}{2} (\bar{\psi}\ga_c)_A e^c ,
\ \bar{\Pi}^{\be A} =\f{\partial L_\D^\be}{\partial d\bar{\psi}_A} =-\f{1-\be}{2} e^c (\ga_c\psi)^A. \la{def_Pi}
\eea
Corresponding conjugate fields $\pi_{(\be)A}^{\mu}$ and $\bar{\pi}_{(\be)}^{A \mu}$ are given by 
\bea
\pi_{(\be)A}^{\mu} =-\f{1+\be}{2}(\bar{\psi}\ga_c)_A \theta^{c\mu} ,\ \bar{\pi}_{(\be)}^{A \mu}=\f{1-\be}{2} \theta^{c\mu} (\ga_c\psi)^A. 
\eea
$\pi_{(\be)}^0,\cdots,\pi_{(\be)}^3(\bar{\pi}_{(\be)}^0,\cdots,\bar{\pi}_{(\be)}^3)$ are not independent each other and not independent from $\bar{\psi}(\psi)$. 
Then, we need to use the method of Lagrange multiplier. 
The Hamilton form is given by
\bea
H_{\D,\tot}^\be \aeq H_\D^\be+\Big[\Pi^\be_A-\f{1+\be}{2} (\bar{\psi}\ga_c)_A e^c\Big]\w \lm^A+\bar{\lm}_A\w \Big[\bar{\Pi}^{\be A}+\f{1-\be}{2} e^c (\ga_c\psi)^A\Big] ,\la{H_D,tot}
\eea
with 
\bea
 H_\D^\be \aeq d\psi^A \w \Pi^\be_A+d\bar{\psi}_A \w \bar{\Pi}^{\be A}-L_\D^\be \no\\
\aeq \f{1+\be}{2} \bar{\psi}\ga_c e^c \w \f{1}{4}\ga_{ab}\om^{ab}\psi+\f{1-\be}{2} e^c \w \f{1}{4}\bar{\psi}\ga_{ab}\om^{ab}\ga_c\psi
+m\bar{\psi}\psi \Om+\f{\be}{2}C_a\bar{\psi}\ga^a\psi \Om .
\eea
Here, $\lm^A$ and $\bar{\lm}_A$ are the Lagrange multiplier 1-forms. 
While the traditional Hamiltonian density includes $\partial_i \psi$, $\partial_i \bar{\psi}$ $(i=1,2,3)$, the Hamilton form does not include 
the exterior derivative of the Dirac fields.
At $\be=1$, $\bar{\pi}_{(\be)}^\mu$ becomes zero. 
In the traditional analytic mechanics at $\be=1$, the treatment which regards only $\psi$ and $\pi_{(1)}^0=-\bar{\psi}\ga_c \theta^{c0}$
as the arguments of the Hamiltonian density goes well.
However, in the CAM and the DW theory, one need to use the Lagrange multiplier even at $\be=1$. 
If we regard $\Pi^{\be=1}$ as independent variable, $d\Pi^{\be=1}_A=-\partial H_\D^{\be=1}/\partial \psi^A$ provides \re{DE2} correctly, 
however, $d\psi^A=-\partial H_\D^{\be=1}/\partial \Pi^{\be=1}_A$ leads incorrect equation\cite{Me}
\bea
d\psi +\f{1}{4}\ga_{ab}\om^{ab}\psi +\f{m}{D} \ga_a\theta^a\psi+\half  C_a  \theta^a \psi = 0 .
\eea
Applying $\ga_b e^b$ to the above equation from the left and using $e^b\w \theta^a=\eta^{ab}\Om$, we obtain \re{DE}. 
From \re{H_D,tot}, we obtain 
\bea
\f{\partial H_{\D,\tot}^\be}{\partial \Pi^\be} \aeq \lm ,\ \f{\partial H_{\D,\tot}^\be}{\partial \bar{\Pi}^\be} = -\bar{\lm},\\
\f{\partial H_{\D,\tot}^\be}{\partial \psi} \aeq \f{1+\be}{2} \bar{\psi}\ga_c e^c \w \f{1}{4}\ga_{ab}\om^{ab}+\f{1-\be}{2} e^c \w \f{1}{4}\bar{\psi}\ga_{ab}\om^{ab}\ga_c
+m\bar{\psi} \Om\no\\
&& +\f{\be}{2}C_a\bar{\psi}\ga^a \Om +\bar{\lm}\w \f{1-\be}{2} e^c \ga_c \la{der_Htot1} ,\\
\f{\partial H_{\D,\tot}^\be}{\partial \bar{\psi}} \aeq \f{1+\be}{2} \ga_c e^c \w \f{1}{4}\ga_{ab}\om^{ab}\psi+\f{1-\be}{2} e^c \w \f{1}{4}\ga_{ab}\om^{ab}\ga_c\psi
+m\psi \Om\no\\
&& +\f{\be}{2}C_a\ga^a\psi \Om-\f{1+\be}{2} \ga_c e^c\w \lm . \la{der_Htot2}
\eea
Then, the canonical equations $d\psi=-\partial H_{\D,\tot}^\be/\partial \Pi^\be$ and $d\bar{\psi}=-\partial H_{\D,\tot}^\be/\partial \bar{\Pi}^\be$ 
respectively become 
\bea
d\psi \aeq -\lm,\ d\bar{\psi} = \bar{\lm} .\la{C_lm}
\eea
The canonical equations $d\Pi^\be=-\partial H_{\D,\tot}^\be/\partial \psi$ and $d\bar{\Pi}^\be=-\partial H_{\D,\tot}^\be/\partial \bar{\psi}$ respectively become 
\bea
d\Pi^\be \aeq  \f{1}{4} \bar{\psi}\ga_{ab}\om^{ab}\w e^c \ga_c-\f{1+\be}{2}\bar{\psi}\ga^c \om_c \Om
-m\bar{\psi}\Om-\f{\be}{2}C_a\bar{\psi}\ga^a\Om -\f{1-\be}{2} d\bar{\psi} \w e^c \ga_c , \la{D1} \\
d\bar{\Pi}^\be \aeq- \ga_c e^c \w \f{1}{4}\ga_{ab}\om^{ab}\psi+\f{1-\be}{2} \ga^c \om_c \psi \Om
-m\psi \Om-\f{\be}{2}C_a\ga^a\psi \Om -\f{1+\be}{2} \ga_c e^c\w d\psi,  \la{D3}
\eea
using \re{der_Htot1}, \re{der_Htot2}, the second equation of \re{trick} and \re{C_lm}.
From \re{def_Pi} and the first equation of \re{trick}, the left hand sides of the above two equations are given by
\bea
 d\Pi^\be \aeq \f{1+\be}{2} d\bar{\psi}\ga_c \w e^c-\f{1+\be}{2} \bar{\psi}\ga^c (\om_c+C_c)\Om \la{dPi},\\
 d\bar{\Pi}^\be \aeq \f{1-\be}{2} (\om_c+C_c) \ga^c\psi \Om+\f{1-\be}{2} e^c \ga_c\w d\psi . \la{dbarPi}
\eea
Substituting these into \re{D1} and \re{D3}, we obtain \re{DE2} and \re{DE} respectively.

We confirm the expectation mentioned at the last of \res{Relation}.
We take only $\psi^A$ and $\bar{\psi}_A$ as the independent variables. 
In contrast to the traditional analytic mechanics, the conjugate forms are the dependent variables. 
The variation of $\Pi^\be_A$ and $\bar{\Pi}^{\be A}$ are respectively given by 
$\dl \bar{\psi}_B \f{\partial \Pi^\be_A}{\partial \bar{\psi}_B}$ and $ \f{\partial \bar{\Pi}^{\be A}}{\partial \psi^B}\dl \psi^B$. 
Then, the variation of the Hamilton form $H_\D^\be(\psi,\bar{\psi}) = d\psi^A \w \Pi^\be_A+d\bar{\psi}_A \w \bar{\Pi}^{\be A}-L_\D^\be$ is given by
\bea
\dl H_\D^\be(\psi,\bar{\psi}) \aeq \dl\psi^A \Big[ d\bar{\psi}_B \w  \f{\partial \bar{\Pi}^{\be B}}{\partial \psi^A} - \f{\partial L_\D^\be}{\partial \psi^A}\Big]+
\dl \bar{\psi}_A \Big[ - \f{\partial \Pi^\be_B}{\partial \bar{\psi}_A}\w d\psi^B -\f{\partial L_\D^\be}{\partial \bar{\psi}_A} \Big].
\eea
From this equation, we obtain
\bea
  d\Pi^\be_A  \aeq -\f{\partial  H_\D^\be}{\partial \psi^A} + d\bar{\psi}_B \w  \f{\partial \bar{\Pi}^{\be B}}{\partial \psi^A},\la{O1}\\
 d\bar{\Pi}^{\be A} \aeq - \f{\partial  H_\D^\be}{\partial \bar{\psi}_A}- \f{\partial \Pi^\be_B}{\partial \bar{\psi}_A}\w d\psi^B \la{O2} ,
\eea
using the Euler-Lagrange equations. We call this method the modified Hamilton formalism. 
$\f{\partial  H_\D^\be}{\partial \psi^A}$ and $\f{\partial  H_\D^\be}{\partial \bar{\psi}_A}$ are respectively $\bar{\lm}\to 0$, $\lm \to 0$ limits of \re{der_Htot1} and \re{der_Htot2}.
The second terms of the right hand sides of \re{O1} and \re{O2} are given by
\bea 
d\bar{\psi}_B \w \f{\partial \bar{\Pi}^{\be B}}{\partial \psi^A }= -\f{1-\be}{2}(d\bar{\psi} \w e^c \ga_c)_A, \
 -\f{\partial \Pi^\be_B}{\partial \bar{\psi}_A}\w d\psi^B  = - \f{1+\be}{2}(\ga_c e^c \w d\psi)^A .
\eea
Then, the right hand sides of \re{O1},\re{O2} are given by 
\bea
-\f{\partial  H_\D^\be}{\partial \psi^A} + d\bar{\psi}_B \w  \f{\partial \bar{\Pi}^{\be B}}{\partial \psi^A} \aeq
 -\f{\partial  H_{\D,\tot}^\be}{\partial \psi^A}\Bv{\bar{\lm}=d\bar{\psi}} ,\\
- \f{\partial  H_\D^\be}{\partial \bar{\psi}_A}- \f{\partial \Pi^\be_B}{\partial \bar{\psi}_A}\w d\psi^B \aeq -\f{\partial  H_{\D,\tot}^\be}{\partial \bar{\psi}_A}\Bv{\lm=-d\psi}. 
\eea
Therefore, \re{O1} and \re{O2} are respectively equivalent to \re{D1} and \re{D3}. 
For the Dirac field, we confirmed that one can obtain correct equation of motions from the Hamilton form by the variation using only independent variables.

\section{Summary}

We showed that the covariant analytic mechanics (CAM) is closely related to the De Donder-Weyl (DW) theory.
Because the DW theory does not consider the constraint that the conjugate fields are complete anti-symmetric tensors, one of 
the DW equations is not correct generally.
Decreasing independent variables using this constraint, we obtained an improved DW equation (\res{Relation}). 
By rewriting the canonical equations of the CAM using the components of the tensors,
we showed that these are equivalent to the improved DW equations (\res{Relation}). 
So, we showed that the CAM is equivalent to the improved DW theory for the first time. 
While the Poisson bracket of the DW theory is defined using the components of the tensors and may describe the incorrect DW equation, 
our Poisson brackets are defined using the differential forms and do not lead to the incorrect DW equation. 
To generalize our Poisson bracket to the Dirac bracket and to apply it to the Dirac field are future works.

In \res{Dirac field}, we applied the CAM to the Dirac field, which is a constraint system.
We treated the Dirac field with Lagrange multipliers.
Getting a hint from the relation between the CAM and the DW theory, 
we presented the modified Hamilton formalism which regards only the Dirac fields as the basic variables and showed it provides the Dirac equations correctly. 

\acknowledgments

We acknowledge helpful discussions with Y. Tokura, S. Tanimura and I. V. Kanatchikov.

\end{document}